\documentclass[11pt]{article}
\usepackage{amsmath,amsthm,amsfonts,comment,array}

\DeclareMathOperator{\Hom}{Hom}
\DeclareMathOperator{\Com}{Com}

\DeclareMathOperator{\1}{id}

\newcommand{\NN}{\mathbb{N}}
\newcommand{\RR}{\mathbb{R}}

\newcommand{\EEnd}{\mathcal End}
\newcommand{\EE}{\mathcal E}
\newcommand{\bul}{\bullet}

\renewcommand{\=}{:=}
\renewcommand{\t}{\otimes}
\renewcommand{\:}{\colon}

\newcommand{\m}{\overset{\circ}{\mu}}
\newcommand{\A}{\hat{A}}
\newcommand{\pp}{\hat{p}}
\newcommand{\q}{\hat{q}}
\newcommand{\muu}{\hat{\mu}}
\newcommand{\qxi}{\hat{\xi}}

\newtheorem{thm}{Theorem}[section]

 \newtheorem{lemma}[thm]{Lemma}

\theoremstyle{definition}
 \newtheorem{defn}[thm]{Definition}

\theoremstyle{definition}

\theoremstyle{definition}
 \newtheorem{rem}[thm]{Remark}

\numberwithin{equation}{section}
\numberwithin{table}{section}

\begin{document}
\title{\LARGE\bf  Quantum counterparts of 3d real Lie algebras\\
 over harmonic oscillator}
\author{\Large Eugen Paal and J\"{u}ri Virkepu 
}
\date{}
\maketitle
\thispagestyle{empty}
\begin{abstract}
Operadic Lax representations for the harmonic oscillator are used to construct the quantum counterparts of 3d real Lie algebras in Bianchi classification. The Jacobians of the quantum algebras are calculated.
\end{abstract}

\section{Introduction}

In Hamiltonian formalism, a mechanical system is described by the
canonical variables $q^i,p_i$ and their time evolution is prescribed
by the Hamiltonian equations
\begin{equation}
\label{ham} \dfrac{dq^i}{dt}=\dfrac{\partial H}{\partial p_i}, \quad
\dfrac{dp_i}{dt}=-\dfrac{\partial H}{\partial q^i}
\end{equation}
By a Lax representation \cite{Lax68,BBT03} of a mechanical system
one means such a pair $(L,M)$ of matrices (linear operators) $L,M$
that the above Hamiltonian system may be represented as the Lax
equation
\begin{equation}
\label{lax} \dfrac{dL}{dt}= ML-LM
\end{equation}
Thus, from the algebraic point of view, mechanical systems can be
represented by linear operators, i.e by  linear maps $V\to V$  of a
vector space $V$. As a generalization of this one can pose the
following question \cite{Paal07}: how to describe the time evolution
of the linear operations (multiplications) $V^{\t n}\to V$?

The algebraic operations (multiplications) can be seen as an example
of the \emph{operadic} variables \cite{Ger}. If an operadic system
depends on time one can speak about \emph{operadic dynamics}
\cite{Paal07}. The latter may be introduced by simple and natural
analogy with the Hamiltonian dynamics. In particular, the time
evolution of the operadic variables may be given by the operadic Lax
equation. In \cite{PV07,PV08,PV08-1}, the low-dimensional binary operadic Lax
representations for the harmonic oscillator were constructed.

In the present paper, the operadic Lax representations for the  harmonic oscillator are used to construct the quantum counterparts of 3d real Lie algebras in Bianchi classification. The Jacobians of the quantum  algebras are calculated.

\section{Endomorphism operad and Gerstenhaber brackets}

Let $K$ be a unital associative commutative ring, $V$ be a unital
$K$-module, and $\EE_V^n\= {\EEnd}_V^n\= \Hom(V^{\t n},V)$
($n\in\NN$). For an \emph{operation} $f\in\EE^n_V$, we refer to $n$
as the \emph{degree} of $f$ and often write (when it does not cause
confusion) $f$ instead of $\deg f$. For example, $(-1)^f\= (-1)^n$,
$\EE^f_V\=\EE^n_V$ and $\circ_f\= \circ_n$. Also, it is convenient
to use the \emph{reduced} degree $|f|\= n-1$. Throughout this paper,
we assume that $\t\= \t_K$.

\begin{defn}[endomorphism operad \cite{Ger}]
\label{HG} For $f\t g\in\EE_V^f\t\EE_V^g$ define the \emph{partial
compositions}
\[
f\circ_i g\= (-1)^{i|g|}f\circ(\1_V^{\t i}\t g\t\1_V^{\t(|f|-i)})
\quad \in\EE^{f+|g|}_V,
         \quad 0\leq i\leq |f|
\]
The sequence $\EE_V\= \{\EE_V^n\}_{n\in\NN}$, equipped with the
partial compositions $\circ_i$, is called the \emph{endomorphism
operad} of $V$.
\end{defn}

\begin{defn}[total composition \cite{Ger}]
The \emph{total composition}
$\bul\:\EE^f_V\t\EE^g_V\to\EE^{f+|g|}_V$ is defined by
\[
f\bul g\= \sum_{i=0}^{|f|}f\circ_i g\quad \in \EE_V^{f+|g|}, \quad |\bul|=0
\]
The pair $\Com\EE_V\= \{\EE_V,\bul\}$ is called the \emph{composition
algebra} of $\EE_V$.
\end{defn}

\begin{defn}[Gerstenhaber brackets \cite{Ger}]
The  \emph{Gerstenhaber brackets} $[\cdot,\cdot]$ are defined in
$\Com\EE_V$ as a graded commutator by
\[
[f,g]\= f\bul g-(-1)^{|f||g|}g\bul f=-(-1)^{|f||g|}[g,f],\quad
|[\cdot,\cdot]|=0
\]
\end{defn}

The \emph{commutator algebra} of $\Com \EE_V$ is denoted as
$\Com^{-}\!\EE_V\= \{\EE_V,[\cdot,\cdot]\}$. One can prove (e.g
\cite{Ger}) that $\Com^-\!\EE_V$ is a \emph{graded Lie algebra}. The
Jacobi identity reads
\[
(-1)^{|f||h|}[[f,g],h]+(-1)^{|g||f|}[[g,h],f]+(-1)^{|h||g|}[[h,f],g]=0
\]

\section{Operadic Lax equation and harmonic oscillator}

Assume that $K\= \RR$ or $K\= \mathbb{C}$ and operations are
differentiable. Dynamics in operadic systems (operadic dynamics) may
be introduced by

\begin{defn}[operadic Lax pair \cite{Paal07}]
Allow a classical dynamical system to be described by the
Hamiltonian system \eqref{ham}. An \emph{operadic Lax pair} is a
pair $(L,M)$ of operations $L,M\in\EE_V$, such that the
Hamiltonian system  (\ref{ham}) may be represented as the
\emph{operadic Lax equation}
\[
\frac{dL}{dt}=[M,L]\= M\bul L-(-1)^{|M||L|}L\bul M
\]
The pair $(L,M)$ is also called an \emph{operadic Lax representations} of/for Hamiltonian system \eqref{ham}.
Evidently, the degree constraints $|M|=|L|=0$ give rise to ordinary
Lax equation (\ref{lax}) \cite{Lax68,BBT03}. In this paper we assume that $|M|=0$.
\end{defn}

The Hamiltonian of the harmonic oscillator (HO) is
\[
H(q,p)=\frac{1}{2}(p^2+\omega^2q^2)
\]
Thus, the Hamiltonian system of HO reads
\begin{equation}
\label{eq:h-osc} \frac{dq}{dt}=\frac{\partial H}{\partial p}=p,\quad
\frac{dp}{dt}=-\frac{\partial H}{\partial q}=-\omega^2q
\end{equation}
If $\mu$ is a linear algebraic operation we can use the above
Hamilton equations to obtain
\[
\dfrac{d\mu}{dt} =\dfrac{\partial\mu}{\partial
q}\dfrac{dq}{dt}+\dfrac{\partial\mu}{\partial p}\dfrac{dp}{dt}
=p\dfrac{\partial\mu}{\partial
q}-\omega^2q\dfrac{\partial\mu}{\partial p}
 =[M,\mu]
\]
Therefore, we get the following linear partial differential equation
for $\mu(q,p)$:
\begin{equation}
\label{eq:diff}
p\dfrac{\partial\mu}{\partial
q}-\omega^2q\dfrac{\partial\mu}{\partial p}=[M,\mu]
\end{equation}
By integrating \eqref{eq:diff} one can get sequences of operations called the
\emph{operadic (Lax representations for) harmonic oscillator}. Since the general solution of a partial differential equation depends on arbitrary functions, these representations are not uniquely determined.

\section{3D binary anti-commutative operadic Lax representations for harmonic oscillator}

\begin{lemma}
\label{lemma:harmonic3} Matrices
\[
L\=\begin{pmatrix}
    p & \omega q & 0 \\
    \omega q & -p & 0 \\
    0 & 0 & 1 \\
  \end{pmatrix},\quad
M\=\frac{\omega}{2}
\begin{pmatrix}
    0 & -1 &0\\
1 & 0 & 0\\
0 & 0 & 0
  \end{pmatrix}
\]
give a 3-dimensional Lax representation for the harmonic oscillator.
\end{lemma}

\begin{defn}[quasi-canonical coordinates]
Denoting by $H$ the Hamiltonian of the harmonic oscillator define its \emph{quasi-canonical coordinates} $A_\pm$ by the relations
\begin{equation}
\label{eq:def_A}
A_+^2+A_-^2=2\sqrt{2H},\quad
A_+^2-A_-^2=2p,\quad
A_+A_-=\omega q
\end{equation}
Note that $A_\pm$ can not be simultaneously zero. 
\end{defn}

\begin{thm}[\cite{PV08-1}]
\label{thm:main}
Let $C_{\nu}\in\mathbb{R}$ ($\nu=1,\ldots,9$) be
arbitrary real--valued parameters, such that
\begin{equation}
\label{eq:cond} C_2^2+C_3^2+C_5^2+C_6^2+C_7^2+C_8^2\neq0
\end{equation}
Let $M$ be defined as in Lemma \ref{lemma:harmonic3}, and
\begin{equation}\label{eq:theorem}
\begin{cases}
\mu_{11}^{1}=\mu_{22}^{1}=\mu_{33}^{1}=\mu_{11}^{2}=\mu_{22}^{2}=\mu_{33}^{2}=\mu_{11}^{3}=\mu_{22}^{3}=\mu_{33}^{3}=0\\
\mu_{23}^{1}=-\mu_{32}^{1}=C_2p-C_3\omega q-C_4\\
\mu_{13}^{2}=-\mu_{31}^{2}=C_2p-C_3\omega q+C_4\\
\mu_{31}^{1}=-\mu_{13}^{1}=C_2\omega q+C_3p-C_1\\
\mu_{23}^{2}=-\mu_{32}^{2}=C_2\omega q+C_3p+C_1\\
\mu_{12}^{1}=-\mu_{21}^{1}=C_5A_++C_6A_-\\
\mu_{12}^{2}=-\mu_{21}^{2}=C_5A_--C_6A_+\\
\mu_{13}^{3}=-\mu_{31}^{3}=C_7A_++C_8A_-\\
\mu_{23}^{3}=-\mu_{32}^{3}=C_7A_--C_8A_+\\
\mu_{12}^{3}=-\mu_{21}^{3}=C_9
\end{cases}
\end{equation}
Then $(\mu,M)$ is a $3$-dimensional anti-commutative binary operadic
Lax pair for HO.
\end{thm}

\section{Initial conditions and dynamical deformations}

Specify the coefficients $C_{\nu}$ in Theorem \ref{thm:main} by the
initial conditions
\[
\left. \mu\right|_{t=0}=\m{}_,\quad
\left.p\right|_{t=0}
=p_0,\quad \left. q\right|_{t=0}=0
\]
Denotinf $E\=H|_{t=0}$, the latter together with \eqref{eq:def_A} yield the initial
conditions for $A_{\pm}$:
\[
\begin{cases}
\left.\left(A_+^{2}+A_-^{2}\right)\right|_{t=0}=2\sqrt{2E}\\
\left.\left(A_+^{2}-A_-^{2}\right)\right|_{t=0}=2p_0\\
\left.A_+A_-\right|_{t=0}=0
\end{cases}
\quad \Longleftrightarrow \quad
\begin{cases}
p_0>0\\
\left.A^{2}_+\right|_{t=0}=2p_0\\
\left.A_-\right|_{t=0}=0
\end{cases}
\vee\quad
\begin{cases}
p_0<0\\
\left.A_+\right|_{t=0}=0\\
\left.A^2_-\right|_{t=0}=-2p_0
\end{cases}
\]
In what follows assume that $p_0>0$ and $A_+|_{t=0}=\sqrt{2p_0}$. The other cases
can be treated similarly. Note that in this case $p_0=\sqrt{2E}$.

From \eqref{eq:theorem} we get the following linear system:
\begin{equation}
\label{eq:constants} \left\{
  \begin{array}{lll}
    \m{}_{23}^{1}=C_2p_0-C_4, & \m{}_{31}^{1}=C_3p_0-C_1, & \m{}_{12}^{1}=C_5\sqrt{2p_0}\\
    \m{}_{13}^{2}=C_2p_0+C_4, &
    \m{}_{12}^{2}=-C_6\sqrt{2p_0}, &
    \m{}_{23}^{2}=C_3p_0+C_1\\
    \m{}_{13}^{3}=C_7\sqrt{2p_0}, &
\m{}_{23}^{3}=-C_8\sqrt{2p_0}, & \m{}_{12}^{3}=C_9
\end{array}
\right.
\end{equation}
One can easily check that the unique solution of the latter system
with respect to $C_\nu$ ($\nu=1,\ldots,9$) is
\[
\left\{
  \begin{array}{lll}
C_1=\frac{1}{2}\left(\overset{\circ}{\mu}{}_{23}^{2}-\overset{\circ}{\mu}{}_{31}^{1}\right),&
C_2=\frac{1}{2p_0}\left(\overset{\circ}{\mu}{}_{13}^{2}+\overset{\circ}{\mu}{}_{23}^{1}\right),&
C_3=\frac{1}{2p_0}\left(\overset{\circ}{\mu}{}_{23}^{2}+\overset{\circ}{\mu}{}_{31}^{1}\right)\vspace{1mm}\\
C_4=\frac{1}{2}\left(\overset{\circ}{\mu}{}_{13}^{2}-\overset{\circ}{\mu}{}_{23}^{1}\right),&
C_5=\frac{1}{\sqrt{2p_0}}\overset{\circ}{\mu}{}_{12}^{1},&
C_6=-\frac{1}{\sqrt{2p_0}}\overset{\circ}{\mu}{}_{12}^{2}\vspace{1mm}\\
C_7=\frac{1}{\sqrt{2p_0}}\overset{\circ}{\mu}{}_{13}^{3},&
C_8=-\frac{1}{\sqrt{2p_0}}\overset{\circ}{\mu}{}_{23}^{3},&
C_9=\overset{\circ}{\mu}{}_{12}^{3}
\end{array}
\right.
\]

\begin{rem}
Note that the parameters $C_{\nu}$  have to satisfy condition \eqref{eq:cond} to get the operadic Lax representations.
\end{rem}

\begin{defn}
If $\mu=\overset{\circ}{\mu}$, then the multiplication $\overset{\circ}{\mu}$ is called \emph{dynamically rigid} (over HO). Otherwise $\mu$ is called a \emph{dynamical deformation} of $\overset{\circ}{\mu}$ (over HO).
\end{defn}

\section{Bianchi classification of 3d real Lie algebras}

We use the Bianchi classification of the 3-dimensional real Lie algebras given in
\cite{Landau80}.
The structure equations of the 3-dimensional real Lie algebras can be presented
as follows:
\[
[e_1,e_2]=-\alpha e_2+n^{3}e_3,\quad
[e_2,e_3]=n^{1}e_1,\quad
[e_3,e_1]=n^{2}e_2+\alpha e_3
\]
The values of the parameters $\alpha,n^{1}, n^{2},n^{3}$  and the corresponding structure constants are presented in Table \ref{table:Bianchi1}.
\begin{table}[ht]
\begin{center}
\begin{tabular}{|c||c||c|c|c||c|c|c|c|c|c|c|c|c|c|c|}\hline
Bianchi type & $\alpha$ & $n^{1}$ & $n^{2}$ & $n^{3}$ &
$\overset{\circ}{\mu}{}_{12}^{1}$ &
$\overset{\circ}{\mu}{}_{12}^{2}$ &
$\overset{\circ}{\mu}{}_{12}^{3}$ &
 $\overset{\circ}{\mu}{}_{23}^{1}$ & $\overset{\circ}{\mu}{}_{23}^{2}$ & $\overset{\circ}{\mu}{}_{23}^{3}$
  & $\overset{\circ}{\mu}{}_{31}^{1}$ & $\overset{\circ}{\mu}{}_{31}^{2}$ &
  $\overset{\circ}{\mu}{}_{31}^{3}$\\\hline\hline
 I & 0 & 0 & 0 & 0 & 0 & 0 & 0 & 0 & 0 & 0 & 0 & 0 & 0
\\\hline
II & 0 & $1$ & 0 & 0 & 0 & 0 & 0 & $1$ & 0 & 0 & 0 & 0 & 0
\\\hline
VII & 0 & $1$ & $1$ & 0 & 0 & 0 & 0 & $1$ & 0 & 0 & 0 & $1$ & 0
\\\hline
VI & 0 & $1$ & $-1$ & 0 & 0 & 0 & 0 & $1$ & 0 & 0 & 0 & $-1$ & 0
\\\hline
IX & 0 & $1$ & $1$ & $1$ & 0 & 0 & $1$ & $1$ & 0 & 0 & 0 & $1$ & 0
\\\hline
VIII & 0 & $1$ & $1$ & $-1$ & 0 & 0 & $-1$ & $1$ & 0 & 0 & 0 & $1$ &
0
\\\hline
V & 1 & 0 & 0 & 0 & 0 & $-1$ & 0 & 0 & 0 & 0 & 0 & 0 & $1$
\\\hline
IV & 1 & 0 & 0 & 1 & 0 & $-1$ & $1$ & 0 & 0 & 0 & 0 & 0 & $1$
\\\hline
VII$_{a}$ & $a$ & 0 & $1$ & $1$ & 0 & $-a$ & $1$ & 0 & 0 & 0 &
0 & $1$ & $a$
\\\hline
III$_{a=1}$& 1 & 0 & $1$ & $-1$ & 0 & $-1$ & $-1$ & 0 & 0 & 0 & 0 &
$1$ & $1$
\\\hline
VI$_{a\neq 1}$& $a$ & 0 & $1$ & $-1$ & 0 & $-a$ & $-1$ & 0 & 0 & 0 &
0 & $1$ & $a$
\\\hline
\end{tabular}
\end{center}
\caption{3d real Lie algebras in Bianchi classification. Here $a>0$.}
\label{table:Bianchi1}
\end{table}

The Bianchi classification is for instance used (see e.g \cite{Landau80}) to
describe the spatially homogeneous spacetimes of dimension 3+1.

\section{Dynamical deformations of 3d real Lie algebras}

By using the structure constants of the 3-dimensional Lie algebras
in the Bianchi classification, Theorem \ref{thm:main} and relations
\eqref{eq:constants} one can propose that evolution of the 3-dimensional algebras can be prescribed as given in Table \ref{table:Bianchi3}.

\begin{table}[!h]
\begin{center}\setlength\extrarowheight{4pt}
\begin{tabular}{|c||c|c|c|c|c|c|c|c|c|c|c|}\hline
Dynamical Bianchi type & $\mu_{12}^{1}$ & $\mu_{12}^{2}$ &
$\mu_{12}^{3}$ & $\mu_{23}^{1}$ & $\mu_{23}^{2}$ & $\mu_{23}^{3}$ &
$\mu_{31}^{1}$ & $\mu_{31}^{2}$ &  $\mu_{31}^{3}$
\\[1.5ex]\hline\hline
I$^{t}$ & 0 & 0 & 0 & 0 & 0 & 0 & 0 & 0 & 0
\\ [1.5ex] \hline
II$^{t}$ & 0 & 0 & 0 & $\frac{p+p_0}{2p_0}$ & $\frac{\omega
q}{2p_0}$ & 0 & $\frac{\omega q}{2p_0}$ & $\frac{p-p_0}{-2p_0}$ & 0
\\ [1.5ex] \hline
VII$^{t}$ & 0 & 0 & 0 & $1$ & 0 & 0 & 0 & $1$ & 0
\\ [1.5ex] \hline
VI$^{t}$ & 0 & 0 & 0 & $\frac{p}{p_0}$& $\frac{\omega q}{p_0}$ & 0 &
$\frac{\omega q}{p_0}$ & $-\frac{p}{p_0}$ & 0
\\ [1.5ex] \hline
IX$^{t}$ & 0 & 0 & $1$ & $1$ & 0 & 0 & 0 & $1$ & 0
\\ [1.5ex] \hline
VIII$^{t}$ & 0 & 0 & $-1$ & $1$ & 0 & 0 & 0 & $1$ & 0
\\ [1.5ex] \hline
V$^{t}$ & $\frac{A_-}{\sqrt{2p_0}}$ & $\frac{-A_+}{\sqrt{2p_0}}$ & 0
& 0 & 0 & $\frac{-A_-}{\sqrt{2p_0}}$ & 0 & 0 &
$\frac{A_+}{\sqrt{2p_0}}$
\\ [1.5ex] \hline
IV$^{t}$ & $\frac{A_-}{\sqrt{2p_0}}$ & $\frac{-A_+}{\sqrt{2p_0}}$ &
$1$ & 0 & 0 & $\frac{-A_-}{\sqrt{2p_0}}$ & 0 & 0 &
$\frac{A_+}{\sqrt{2p_0}}$
\\ [1.5ex] \hline
VII$^{t}_a$ & $\frac{aA_-}{\sqrt{2p_0}}$ &
$\frac{-aA_+}{\sqrt{2p_0}}$ & $1$ & $\frac{p-p_0}{-2p_0}$ &
$\frac{\omega q}{-2p_0}$ & $\frac{-aA_-}{\sqrt{2p_0}}$ &
$\frac{\omega q}{-2p_0}$ & $\frac{p+p_0}{2p_0}$ &
$\frac{aA_+}{\sqrt{2p_0}}$
\\ [1.5ex] \hline
III$_{a=1}^{t}$ & $\frac{A_-}{\sqrt{2p_0}}$ &
$\frac{-A_+}{\sqrt{2p_0}}$ & $-1$ & $\frac{p-p_0}{-2p_0}$ &
$\frac{\omega q}{-2p_0}$ & $\frac{-A_-}{\sqrt{2p_0}}$ & $\frac{\omega
q}{-2p_0}$ & $\frac{p+p_0}{2p_0}$ & $\frac{A_+}{\sqrt{2p_0}}$
\\ [1.5ex] \hline
VI$_{a\neq1}^{t}$ & $\frac{aA_-}{\sqrt{2p_0}}$ &
$\frac{-aA_+}{\sqrt{2p_0}}$ & $-1$ & $\frac{p-p_0}{-2p_0}$ &
$\frac{\omega q}{-2p_0}$ & $\frac{-aA_-}{\sqrt{2p_0}}$ &
$\frac{\omega q}{-2p_0}$ & $\frac{p+p_0}{2p_0}$ &
$\frac{aA_+}{\sqrt{2p_0}}$
\\ [1.5ex] \hline
\end{tabular}
\end{center}
\caption{Evolution of 3d real Lie algebras. Here $p_0=\sqrt{2E}$.}
\label{table:Bianchi3}
\end{table}

\begin{thm}[dynamically rigid algebras]
The algebras I, VII, VIII, and IX are dynamically rigid over the harmonic oscillator.
\end{thm}
\begin{proof}[Proof]
This is evident from Tables \ref{table:Bianchi1} and \ref{table:Bianchi3}.
\end{proof}

\begin{thm}[dynamical Lie algebras \cite{PV08-2}]
\label{thm:lie}
The algebras II$\,^{t}$, IV$\,^{t}$, V$\,^{t}$, VI$\,^{t}$, III$_{a=1}^{\,t}$, VI$_{a\neq1}^{\,t}$, and VII$^{\,t}_a$ are Lie algebras.
\end{thm}

Thus, we can see that evolution of these algebras are generated by the harmonic oscillator, because their multiplications depend on the canonical and quasi-canonical coordinates of the harmonic oscillator.

\section{Quantum algebras from the Bianchi classification}

By using the dynamically deformed Bianchi classification (see Table \ref{table:Bianchi3}), we can now propose the canonically quantized counterparts of the 3d Lie algebras in the Bianchi classification (see Table \ref{table:Bianchi4}).
We shall fix the meaqning of the parameter $p_0\=\sqrt{2E}>0$ later.

\begin{table}[!h]
\begin{center}\setlength\extrarowheight{4pt}
\begin{tabular}{|c||c|c|c|c|c|c|c|c|c|c|c|}\hline
Quantum Bianchi type & $\muu_{12}^{1}$ & $\muu_{12}^{2}$ &
$\muu_{12}^{3}$ & $\muu_{23}^{1}$ & $\muu_{23}^{2}$ &
$\muu_{23}^{3}$ & $\muu_{31}^{1}$ & $\muu_{31}^{2}$ &
$\muu_{31}^{3}$
\\[1.5ex]\hline\hline
I$^{\hbar}$ & 0 & 0 & 0 & 0 & 0 & 0 & 0 & 0 & 0
\\ [1.5ex] \hline
II$^{\hbar}$ & 0 & 0 & 0 & $\frac{\pp+p_0}{2p_0}$ & $\frac{\omega
\q}{2p_0}$ & 0 & $\frac{\omega \q}{2p_0}$ & $\frac{\pp-p_0}{-2p_0}$ & 0
\\ [1.5ex] \hline
VII$^{\hbar}$ & 0 & 0 & 0 & $1$ & 0 & 0 & 0 & $1$ & 0
\\ [1.5ex] \hline
VI$^{\hbar}$ & 0 & 0 & 0 & $\frac{\pp}{p_0}$& $\frac{\omega \q}{p_0}$ & 0 &
$\frac{\omega \q}{p_0}$ & $-\frac{\pp}{p_0}$ & 0
\\ [1.5ex] \hline
IX$^{\hbar}$ & 0 & 0 & $1$ & $1$ & 0 & 0 & 0 & $1$ & 0
\\ [1.5ex] \hline
VIII$^{\hbar}$ & 0 & 0 & $-1$ & $1$ & 0 & 0 & 0 & $1$ & 0
\\ [1.5ex] \hline
V$^{\hbar}$ & $\frac{\A_-}{\sqrt{2p_0}}$ & $\frac{-\A_+}{\sqrt{2p_0}}$ & 0
& 0 & 0 & $\frac{-\A_-}{\sqrt{2p_0}}$ & 0 & 0 &
$\frac{\A_+}{\sqrt{2p_0}}$
\\ [1.5ex] \hline
IV$^{\hbar}$ & $\frac{\A_-}{\sqrt{2p_0}}$ & $\frac{-\A_+}{\sqrt{2p_0}}$ &
$1$ & 0 & 0 & $\frac{-\A_-}{\sqrt{2p_0}}$ & 0 & 0 &
$\frac{\A_+}{\sqrt{2p_0}}$
\\ [1.5ex] \hline
VII$^{\hbar}_a$ & $\frac{a\A_-}{\sqrt{2p_0}}$ &
$\frac{-a\A_+}{\sqrt{2p_0}}$ & $1$ & $\frac{\pp-p_0}{-2p_0}$ &
$\frac{\omega \q}{-2p_0}$ & $\frac{-a\A_-}{\sqrt{2p_0}}$ &
$\frac{\omega \q}{-2p_0}$ & $\frac{\pp+p_0}{2p_0}$ &
$\frac{a\A_+}{\sqrt{2p_0}}$
\\ [1.5ex] \hline
III$_{a=1}^{\hbar}$ & $\frac{\A_-}{\sqrt{2p_0}}$ &
$\frac{-\A_+}{\sqrt{2p_0}}$ & $-1$ & $\frac{\pp-p_0}{-2p_0}$ &
$\frac{\omega \q}{-2p_0}$ & $\frac{-\A_-}{\sqrt{2p_0}}$ & $\frac{\omega
\q}{-2p_0}$ & $\frac{\pp+p_0}{2p_0}$ & $\frac{\A_+}{\sqrt{2p_0}}$
\\ [1.5ex] \hline
VI$_{a\neq1}^{\hbar}$ & $\frac{a\A_-}{\sqrt{2p_0}}$ &
$\frac{-a\A_+}{\sqrt{2p_0}}$ & $-1$ & $\frac{\pp-p_0}{-2p_0}$ &
$\frac{\omega \q}{-2p_0}$ & $\frac{-a\A_-}{\sqrt{2p_0}}$ &
$\frac{\omega \q}{-2p_0}$ & $\frac{\pp+p_0}{2p_0}$ &
$\frac{a\A_+}{\sqrt{2p_0}}$
\\ [1.5ex] \hline
\end{tabular}
\end{center}
\caption{Quantum algebras over the harmonic oscillator. Here $p_0=\sqrt{2E}$.}
\label{table:Bianchi4}
\end{table}

The quantum conditions for $\A_\pm$ were proposed and studied in \cite{PV08-3}. Here we study the Jacobi identities for the quantum algebras from Table \ref{table:Bianchi4}. Denoting $\muu:=[\cdot,\cdot]_\hbar$, the quantum Jacobian is defined by
\begin{align*}
\hat{J}_\hbar(x,y,z)
&:=[[x,y]_\hbar,z]_\hbar+[[y,z]_\hbar,x]_\hbar+[[z,x]_\hbar,y]_\hbar\\
&\,\,=\hat{J}^1_\hbar(x,y,z)e_1+\hat{J}^2_\hbar(x,y,z)e_2+\hat{J}^3_\hbar(x,y,z)e_3
\end{align*}

\begin{thm}[rigid algebras]
I, VII, VIII, and IX are rigid over the quantum harmonic oscillator.
\end{thm}

\begin{proof}
This is evident from Tables \ref{table:Bianchi1} and \ref{table:Bianchi4}.
\end{proof}

\begin{thm}[quantum Lie algebras]
\label{thm:rigid}
 II$\,^{\hbar}$ and VI$\,^{\hbar}$  are Lie algebras.
\end{thm}

\begin{proof}
By direct calculations without using constraints and the (quasi-) CCR we can see that
\[
\hat{J}^{1}_\hbar (x,y,z)=\hat{J}^{2}_\hbar (x,y,z)=\hat{J}^{3}_\hbar (x,y,z)=0
\tag*{\qed}
\]
\renewcommand{\qed}{}
\end{proof}

\begin{thm}[anomalous quantum algebras of the first type]
 IV$\,^{\hbar}$ and V$\,^{\hbar}$  are non-Lie algebras.
\end{thm}

\begin{proof}
By direct calculations one can see that 
\begin{align*}
\hat{J}^{1}_\hbar (x,y,z)
&=0=\hat{J}^{2}_\hbar (x,y,z)\\
\hat{J}^{3}_\hbar(x,y,z)
&=\frac{(x|y|z)}{p_0}[\A_+,\A_-] \tag*{\qed}
\end{align*}
\renewcommand{\qed}{}
\end{proof}

\begin{thm}[anomalous quantum algebras of the second type]
\label{thm:lie}
III$_{a=1}^{\,\hbar}$, VI$_{a\neq1}^{\,\hbar}$, and VII$^{\,\hbar}_a$ are non-Lie algebras.
\end{thm}

\begin{proof}
Denote the scalar triple product of the vectors $x,y,z$ by
\[
(x|y|z)\= 
\begin{vmatrix}
 x^{1} & x^{2} & x^{3} \\
 y^{1} & y^{2} & y^{3} \\
z^{1} & z^{2} & z^{3} \\
\end{vmatrix}
\]
and
\[
\qxi_{\pm} \=\omega\q\A_\mp\pm (\pp \mp p_0)\A_\pm
\]
Then, by direct calculations one can check that for the algebras
VI$_{a\neq1}^{\,\hbar}$ and
VII$^{\,\hbar}_a$ the Jacobian coordinates are
\[
      \hat{J}^{1}_\hbar(x,y,z)=\frac{a\tau(x|y|z)}{\sqrt{2p_0^{3}}}\qxi_{+},\quad
      \hat{J}^{2}_\hbar(x,y,z)=\frac{a\tau(x|y|z)}{\sqrt{2p_0^{3}}}\qxi_{-},\quad
      \hat{J}^{3}_\hbar(x,y,z)=\frac{a^{2}(x|y|z)}{p_0}[\A_+,\A_-]
 \]
where $\tau=-1$ and for the algebra III$_{a=1}^{\hbar}$ one has the same formulae with $a=1$.
\end{proof}

More closely these formulae are studied in \cite{PV08-3}.

\section*{Acknowledgements}

The research was in part supported by the Estonian Science Foundation, Grant ETF-6912. The authors are grateful to S. Hervik and P. Kuusk for discussions about using the Bianchi  classification   in cosmology.

\medskip
\noindent
Department of Mathematics, Tallinn University of Technology\\
Ehitajate tee 5, 19086 Tallinn, Estonia\\
E-mails: eugen.paal@ttu.ee and jvirkepu@staff.ttu.ee

\end{document}